\documentclass[preprint,12pt]{elsarticle}

\usepackage{graphicx}
\usepackage{amsfonts,amssymb,amsbsy,amsmath,amsthm}
\usepackage{url} 
\newcommand{\rel}{\tilde{\rho_0}}
\newcommand{\ini}{\rho_0^*}

 \biboptions{sort&compress}

\journal{arXiv}

\begin{document}

\begin{frontmatter}


\title{Computational simulation of bone remodelling post reverse total shoulder arthroplasty}



\author[cerecam]{H. Liedtke}
\ead{LDTHEL001@myuct.ac.za}
\author[glasgow,cerecam]{A.T. McBride\corref{cor1}}
\ead{andrew.mcbride@glasgow.ac.uk}
\cortext[cor1]{Corresponding author}
\author[biomed]{S. Sivarasu}
\ead{sudesh.sivarasu@uct.ac.za}
\author[orthopod]{S. Roche}
\ead{stephen.roche@uct.ac.za]}

\address[cerecam]{Centre for Research in Computational and Applied Mechanics (CERECAM), University of Cape Town, 7701 Rondebosch, South Africa}
\address[glasgow]{School of Engineering, Division of Infrastructure and Environment, University of Glasgow, G128QQ, United Kingdom}
\address[biomed]{Division of Biomedical Engineering, University of Cape Town,  7701 Rondebosch, South Africa}
\address[orthopod]{Orthopaedics Department, University of Cape Town, 7701 Rondebosch, South Africa}

\begin{abstract}
Bone is a living material. 
It adapts, in an optimal sense, to loading by changing its density and trabeculae architecture - a process termed remodelling. 
Implanted orthopaedic devices can significantly alter the loading on the surrounding bone, which can have a detrimental impact on bone ingrowth that is critical to ensure secure implant fixation. 
In this contribution, a computational model that accounts for bone remodelling is developed to elucidate the response of bone following a reverse shoulder procedure for rotator cuff deficient patients. 
The physical process of remodelling is modelled using continuum scale, open system thermodynamics whereby the density of bone evolves isotropically in response to the loading it experiences. 
The fully-nonlinear continuum theory is solved approximately using the finite element method.
The code developed to model the reverse shoulder procedure is validated using a series of benchmark problems.
\end{abstract}


\end{frontmatter}









\section{Introduction}

\begin{figure}[htb!]
\centering
	\includegraphics[keepaspectratio=true,width=1.0\textwidth]{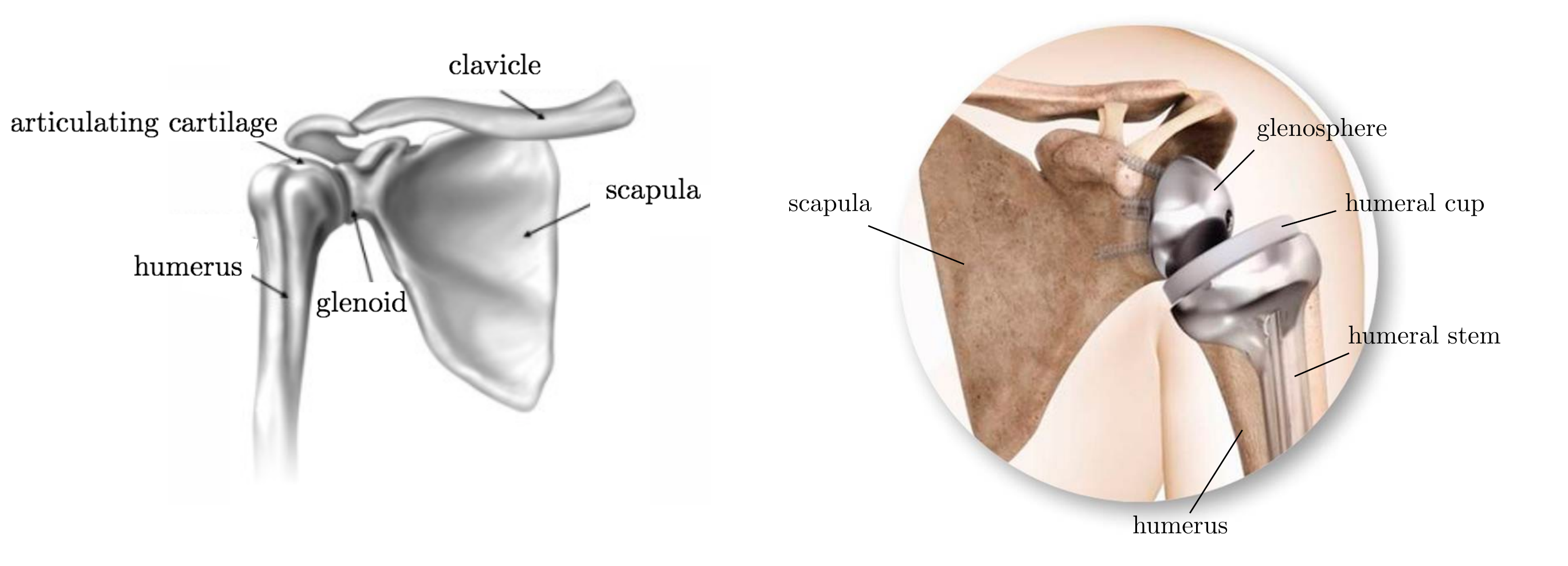}
	\caption{An illustration of the anatomical and reverse shoulder \citep[adapted from][]{shoulder_bones, rtsa2}.}
	\label{fig:rtsa2}
\end{figure}

Reverse total shoulder arthroplasty is a surgical procedure to restore the integrity and functionality of the shoulder joint for rotator cuff deficient patients. 
The rotator cuff muscles stabilise the dynamic shoulder joint. 
In the reverse procedure, the humeral head is removed and a stem is placed in the central channel of the humerus with a polyethylene cup press-fitted into the top of the stem, see Figure~\ref{fig:rtsa2}. 
A hemispherical implant, termed the glenosphere, is screwed into the glenoid component of the scapula (shoulder blade).
The deltoid muscle now compensates for the absent rotator cuff muscles, thereby restoring most of the functionality of the shoulder. 

Even though the procedure restores functionality, many complications are reported post procedure - up to 75 \% in some clinical series \citep{Hsu2011}.
The most common complication is scapular notching \citep[see][]{Hsu2011, Boileau2005, Virani2008, Gutierrez2011, Berliner2015},
where, at full adduction, the humeral cup impinges on the inferior border of the scapula. 
The impingement causes bone resorption in the affected regions. 
Further, the polyethylene debris causes infection and osteolysis \cite{Boileau2005}.
It is also hypothesised that the change in the stress distribution in the scapula post procedure may cause atrophy in the inferior region \cite{PersonalCommunication}.

Computational models are a valuable, predictive tool in orthopaedics and for the design of implants.
The effectiveness of different implant designs can be investigated initially with a computational model and, by following an iterative virtual design process, provide the patient with the optimal prosthesis.

The stress distribution in the bone affects the fixation quality of the implant. 
Bone is a complex, living tissue which responds to changes in the loading environment by altering its density and trabecular orientation, resulting in modified material properties.
This phenomenon is termed bone remodelling.
Bone consists of layers of dense cortical bone and spongy trabecular bone which is composed of strut-like structures surrounded by bone marrow.
Furthermore, bone consists of a solid and a fluid phase. 
The remodelling response occurs in the solid phase and only at low strain rates. 
Remodelling is governed by cells in the solid phase, which deposit (osteoblastic action) and remove (osteoclastic action) material from the bone, resulting in density changes. 
For more details on the structure of bone and bone remodelling see \citep{Taber1995, WeinerWagner1998, Cowin1976a}.

Bone is often modelled as a porous solid.
Because the remodelling response of bone is optimal to the loading environment, objective functions have been proposed to constrain and control the density evolution. 
The difference between a mechanical stimulus, which may be a stress, strain or strain-energy density, and an attractor state, which is the optimal state the bone is driving towards, provides the driving force for the density evolution, \citep[see][for further details on objective function based bone remodelling models]{Cowin1976a, Jacobs1997, Harrigan1996}. 

The open system thermodynamics approach introduced by \citet{Kuhl2003a} is adopted here to describe bone remodelling in the scapula pre and post the reverse shoulder procedure. 
The authors are aware of only six works in the open literature on bone remodelling studies in the scapula \citep{Sharma2009, Sharma2010, Sharma2013, Quental2014, Campoli2013, Campoli2014}. 
In these studies various remodelling theories were adopted that differ from the open system approach.
\citet{Sharma2013}, for example, employ a strain-energy based, isotropic remodelling theory based on \citet{Jacobs1997} together with a linear elastic material model.
The initial geometry was provided by computed tomography (CT) scan data.
The simulation results were reported to be over- and underestimating the actual density values determined from the CT scan.
Similarly, \citet{Campoli2013} used a CT scanned scapula as the model geometry. 
An optimisation based density evolution algorithm \citep{Weinans1992} with a strain-based mechanical stimulus was used.
The muscle loads were obtained from the Delft Shoulder and Elbow model \citep{Delft}.
\citeauthor{Campoli2013} compared the simulation results to the actual density values from the CT scan and reported an error less than 30\%.
\citet{Quental2014} also used a three dimensional, subject-specific scapula as the geometry.
Bone was modeled as a linear elastic, orthotropic, porous material following \citet{Fernandes1999} \citep[see also][]{Folgado2009}. 
The optimisation algorithm involved a strain-based mechanical stimulus. 
The bone microstructure is captured by periodic unit cells accounting for the porosity \citep{Folgado2009}, where the density is dependent on the solid volume fraction of the cells. 
The loads were obtained from a biomechanical model \citep{Quental2014}.
\citeauthor{Quental2014} compared the density to the actual density values from the CT scan and reported an absolute error of below 33\%.

The novel contribution of the work presented here is the application of the fully-nonlinear remodelling theory of \citet{Kuhl2003a} to investigate density changes following reverse total shoulder arthroplasty.
In addition, new insights into the theory are provided via a series of numerical examples.
The validated finite element code developed to perform the numerical examples is available at \cite{Code}.

This manuscript is organised as follows.
The governing and constitutive equations for bone remodelling are detailed in Section \ref{Sec:Equations}. 
The numerical model is introduced in Section \ref{Sec:NumericalModel}.
The theory is then validated with a benchmark problem and different aspects of the model are investigated in Section \ref{Sec:Benchmarks}. 
The validated code is applied to the intact scapula, followed by the application to the reverse shoulder in Section \ref{Sec:Scapula}. 
Here, the ASTM F2028 \citep{ASTM_Glenoid} problem of a glenoid component implanted into a polyurethane foam block bone substitute is investigated.
Finally, the reverse shoulder procedure is analysed. 
A discussion of the numerical results and model features are presented and conclusions are drawn in Section \ref{Sec:DiscussionConclusion}.

\subsection*{Notation}
\label{Sec:Notation}

The scalar product of two second-order tensors $\bf{A}$ and $\bf{B}$ is denoted by $ \bf{A}:\bf{B}$.
The material time derivative of a quantity $ \{ \bullet \}$ is denoted by $D_t \{ \bullet \}$. 
The divergence $\text{Div} \{ \bullet \}$ and gradient $\text{Grad} \{ \bullet \}$ operators are taken with respect to the material placement $\bf{X}$.
As the formulations are based on the open system theory, a quantity $\{ \bullet \}$ may be expressed per unit volume $\{ \bullet \}_0$ or per unit mass $\{ \bullet \}$, both with respect to the material configuration.
The two descriptions are related via the material density $\rho_0$ as $\{ \bullet \}_0 = \rho_0 \{ \bullet \}$.
$\bf{I}$ denotes the second-order identity tensor.
For further details, see \citet{Kuhl2003a}.

\section{Governing and constitutive equations for bone remodelling}
\label{Sec:Equations}

Bone is modelled here as a continuum; i.e.\ is treated as continuous distribution of matter at the macroscopic scale. 
The microstructure of bone is thus not directly accounted for.
A brief summary of the relevant nonlinear continuum mechanics is now given, for additional details see \citet{Holzapfel}, for example.

\begin{figure}[htb!]
\centering
	\includegraphics[keepaspectratio=true,width=1.0\textwidth]{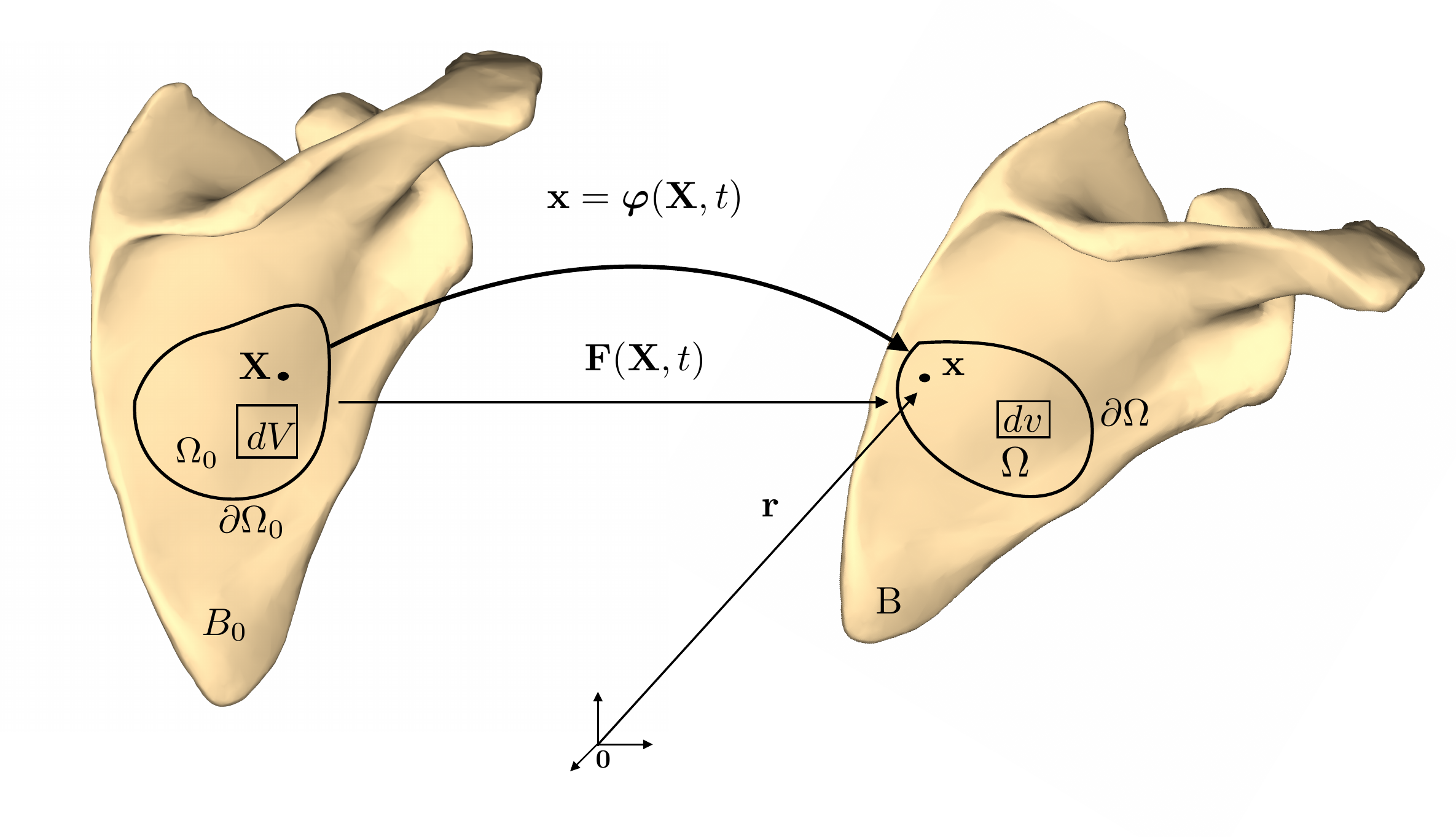}
	\caption{Motion and deformation of a continuum body.}
	\label{fig:continuum}
\end{figure}

Consider the deformation of a body $B_0 \subset \mathbb{R}^3$ from the reference configuration to a current configuration occupied by the deformed body $B$, as shown in Figure \ref{fig:continuum}.
The control regions $\Omega_0 \subset B_0$ and $\Omega \subset B$  with boundaries $\partial \Omega_0$ and $\partial \Omega$, respectively, are shown in Figure \ref{fig:continuum}.
The nonlinear deformation map  $\boldsymbol{\varphi}$, i.e.\ the motion, maps the reference position $\bf{X}$ to the current position $\bf{x}$ at time $t$. 
The gradient of the nonlinear deformation map  $\bf{F} :=  \text{Grad} \boldsymbol{\varphi} $ is termed the deformation gradient.
The determinant of the deformation gradient is defined by $J := \text{det} \bf{F}$ and relates quantities such as the infinitesimal volume elements $dV$ and $dv$ between the two configurations.
The right Cauchy--Green tensor ${\bf{C}} := {\bf{F}}^T {\bf{F}} $ is the chosen measure of deformation.

The theory developed here is based on open system thermodynamics, where the balance relations are formulated with material reference. 
For the sake of brevity, only the final governing equations and constitutive relations are stated. 
For a detailed derivation of the balance relations see \citet{Kuhl2003a}.

The balances of mass and linear momentum are given by 
\begin{align}
D_t \tilde{\rho_0} &= \text{Div} {\bf{R}} + R_0  \, , \label{eq:mass} \\
\text{Div} {\bf{P}} &= {\bf{0}} \, . \label{eq:momentum}
\end{align}
The relative density $\tilde{\rho_0}$, introduced as the primary variable in the balance of mass \eqref{eq:mass}, is defined by
\begin{align*}
\rel = \frac{\rho_0 - \ini}{\ini}
&& \text{where} &&
\begin{matrix} 
 -1 \leq \rel < 0 \text{ resorption } \\
 0 \leq \rel < \infty \text{ absorption }
\end{matrix} \, ,
\end{align*}
and $\ini$ is the initial density (assumed here to be uniform).  
$\bf{R}$ is the mass flux vector and $R_0$ is the scalar mass source term. 

The balance of linear momentum \eqref{eq:momentum} has been expressed in reduced form as the equilibrium equation by neglecting the inertial term. 
The time scales of the inertial and remodelling effects are significantly different, with remodelling taking place over a significantly longer time span.
Body forces are also commonly neglected for clinical problems, as the applied loading during everyday activities is far greater than gravity and is therefore the stimulus for remodelling \cite{Jacobs1997}.

The constitutive relations proposed by \citet{Kuhl2003a} are parameterised by the relative density $\rel$ and the right Cauchy--Green tensor $\bf{C}$. 
Bone is modelled as a compressible neo-Hookean material with a dependence on the density.
The corresponding free energy is given by 
\begin{align}
\Psi(\rel, {\bf{F}}) &= \dfrac{[\rel + 1]^{n-1}}{\ini} \Psi^\text{neo} \label{Psi}
\intertext{with } 
\Psi^\text{neo} &= \dfrac{1}{2} \lambda \text{ln}^2 J + \frac{1}{2} \mu \left[ {\bf{C}} : {\bf{I}} - 3 - 2 \text{ln} J \right] \, ,
\end{align}
where $\lambda$ and $\mu$ are the Lam\'{e} constants, and $n$ is a material parameter which describes the bones porosity and is commonly chosen as 2 \citep{Rice}. 

The constitutive relation for the first Piola--Kirchhoff stress tensor is given by 
\begin{align*}
{\bf{P}} = \ini \left[ \rel + 1 \right] \dfrac{\partial \Psi (\rel, {\bf{F}})}{\partial {\bf{F}}} \, .
\end{align*}
The stress $\bf{P}$ is therefore governed by the coupled influence of the deformation and the density.

The mass flux vector $\bf{R}$ is defined analogous to Fick's law of concentrations or Fourier's law of heat conduction and is given by
\begin{align*}
{\bf{R}} = K_0 \text{Grad} \rel \, ,
\end{align*}
where $K_0\geq 0$ is the mass conduction coefficient. 

The structure of the mass source term $R_0$ is based on the theory developed by \citet{Harrigan1996}.
Following \citet{Kuhl2003a} the mass source term is given by
\begin{align*}
R_0 = \frac{c}{\ini} \left[ \left[ \rel + 1 \right]^{-m} \Psi_0(\rel, {\bf{F}}) - \Psi_0^* \right] \, ,
\end{align*}
where $\Psi_0 = \rho_0 \Psi$.
$c$ governs the speed of the remodelling process and $m$ was introduced by  \citet{HarriganNeuesPaper} to ensure numerical stability. 
$\Psi_0^*$ is the attractor stimulus, which is the reference free energy the system is driving towards. 

\section{Numerical model}
\label{Sec:NumericalModel}

The finite element method is used to approximately solve the nonlinear governing equations developed in the previous sections. 
The displacements and the relative density are introduced as the primary degrees of freedom.
An implicit backward-Euler time stepping scheme is used for the time discretisation. 
The non-linear, coupled system is linearised following a Newton--Raphson approach and solved monolithically.
The finite element code is developed using the AceGen library \cite{Korelc2016}.
The full source code required to reproduce several of the benchmark problems  presented in Section~\ref{Sec:Benchmarks} is available  at \cite{Code}.

\section{Benchmark problems}
\label{Sec:Benchmarks}

In order to validate the code and elucidate features of the formulation, a prototype example of a one-dimensional model problem, set out by \citet{Kuhl2003a}, is presented. 
A further new example investigates the influence of the time step size on the mass flux response.

\subsection{An investigation of the mass source term}

\begin{figure}[htb!]
\centering
	\includegraphics[keepaspectratio=true,width=1.0\textwidth]{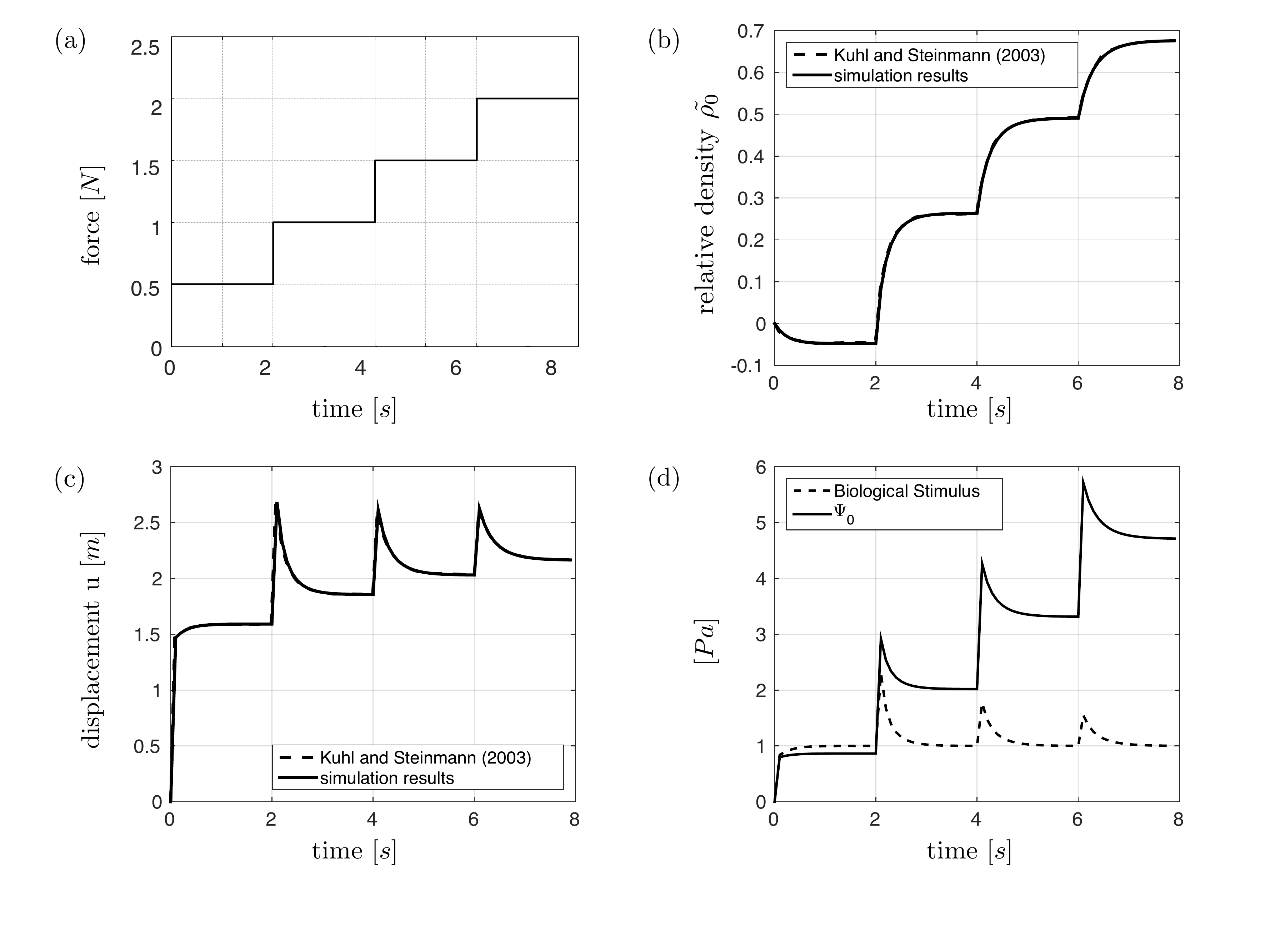}
	\caption{The forcing function is shown in (a). The predicted density and displacement evolution and the results of \citet{Kuhl2003a} are shown in (b) and (c) respectively. The evolution of the free energy $\Psi_0$ and the biological stimulus $\left[ \rel + 1 \right] ^{-m} \Psi_0 $ are shown in (d).}
	\label{fig:KS_Fig1}
\end{figure}

A specimen is loaded in tension by a time-dependent forcing function defined piecewise as shown in Figure \ref{fig:KS_Fig1}(a). 
The specimen has unit length and a cross-sectional of $A = 0.5 \text{m}^2$, a Young's modulus $E = 1 \text{N m}^{-2}$ and a Poisson's ratio $\nu = 0$.
The mass flux $\bf{R}$ is set to zero to investigate the influence of the mass source term.
The initial density $\ini = 1 \text{g cm}^{-3}$ and the attractor stimulus is set to $\Psi_0^* = 1 \text{N m}^{-2}$. 
The exponents $n$ and $m$ are set to $2$ and $3$, respectively.
The time step size is chosen as $\Delta t = 0.1 \text{ s}$.

The evolution of the relative density and the displacement history of a point on the loaded face is shown in Figure \ref{fig:KS_Fig1} (b) and (c), respectively. 
The simulations match exactly the results of \citet{Kuhl2003a}. 
The density and displacement change abruptly as the initial force and the subsequent stepwise increases of the force are applied and then converge to an equilibrium state. 

The evolution of the volume specific, density-weighted free energy $\Psi_0$ and the biological stimulus $\left[ \rel + 1 \right] ^{-m} \Psi_0 $ are shown in Figure \ref{fig:KS_Fig1} (d). The abrupt changes and the subsequent convergence towards an equilibrium state can be observed. 
The coupling between the density, displacement and the free energy is also evident. 
It is important to note that the biological stimulus will always drive towards the attractor stimulus  $\Psi_0^*$, which equals 1 Pa in this example. 
The other quantities are increasing with the increasing force, as expected.

\begin{figure}[htb!]
\centering
	\includegraphics[keepaspectratio=true,width=1.0\textwidth]{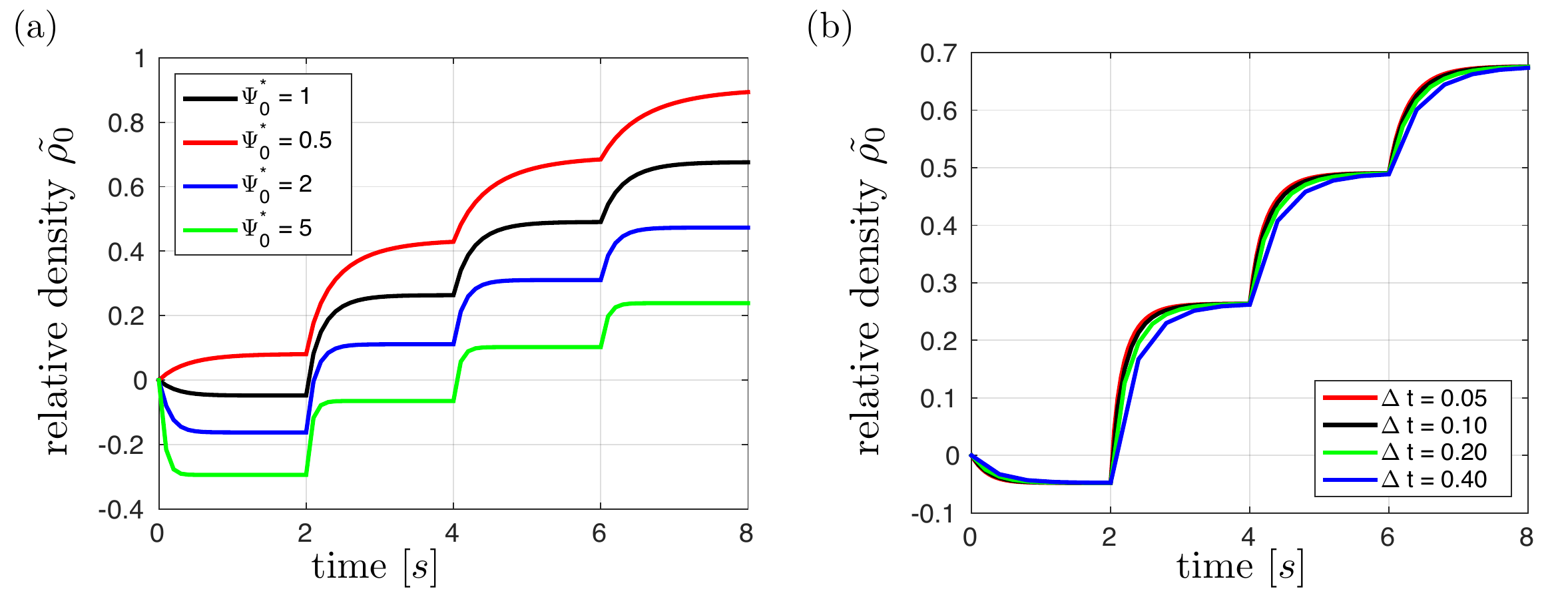}
	\caption{The influence of the attractor stimulus $\Psi_0^*$ and the time step size $\Delta t$ are shown in (a) and (b) respectively.}
	\label{fig:KS_Fig2}
\end{figure}

The influence of the attractor stimulus $\Psi_0^*$ on the density evolution is investigated further, as shown in Figure \ref{fig:KS_Fig2} (a). 
The initial decrease of density observed in \ref{fig:KS_Fig1} (b) is due to the high value of the attractor stimulus. 
There is an inverse relationship between the attractor stimulus and the density. 
This is explained by the nature of the mass source - as the attractor stimulus is smaller, the increase in density is naturally larger.

Finally, the influence of the time step size is investigated, as shown in Figure \ref{fig:KS_Fig2} (b).  
The time step size does not influence the density evolution significantly. 
The convergence rate is affected, but the steady state solution is essentially unchanged.  

\subsection{Investigation of the influence of the time step size and mass flux}

\begin{figure}[htb!]
\centering
	\includegraphics[keepaspectratio=true,width=1.0\textwidth]{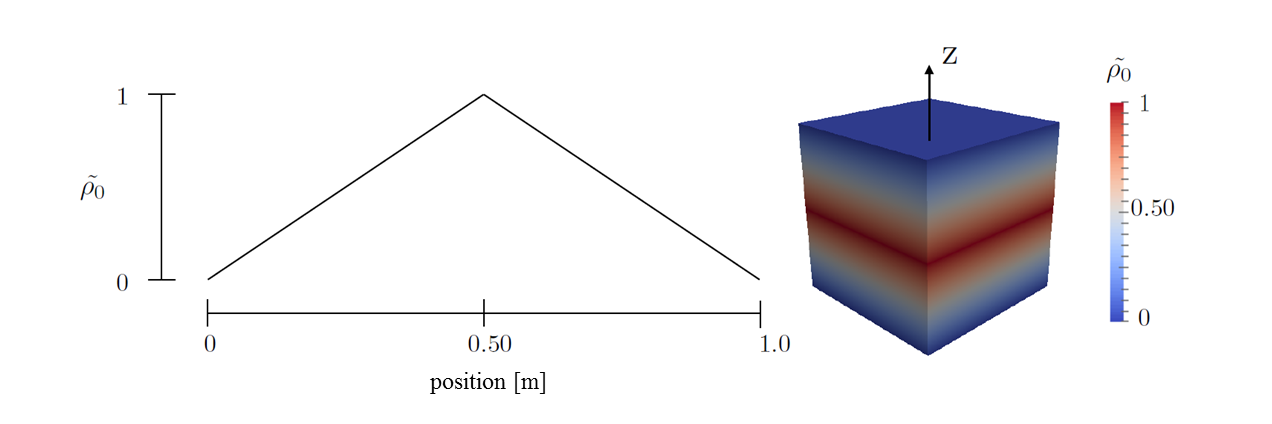}
	\caption{The initial non-uniform density distribution along the length of a unit size, three-dimensional specimen.}
	\label{fig:Flux}
\end{figure}

The choice of time step size becomes important if a mass flux term is introduced. 
Consider now  a unit size, three-dimensional specimen with a non-uniform initial density distribution, as shown in Figure \ref{fig:Flux}.
No mass source or mechanical loading is applied to the system. 
The density evolution is governed by the mass flux alone, i.e.\ \eqref{eq:mass} reduces to $D_t \rel = \text{Div} {\bf{R}}$.
The mass flux term acts to smooth the solution, producing a uniform distribution of $\rel = 0.5$ after 0.4~s if a time step size of $\Delta t = 0.1$~s is used. 
However, for a smaller time step size of $\Delta t = 0.01$~s, equilibrium is reached after 0.18~s.
For a larger time step size of $\Delta t  =1 $~s, equilibrium is reached after 2~s. 
The same behaviour is observed for the problem of linear heat equation in a rigid conductor, where Fourier's law governs the conduction.
This is expected from the nature of the parabolic governing relations, but has not been commented on before in the context of bone remodelling. 

\section{Numerical prediction of remodelling in the scapula pre and post reverse shoulder procedure}
\label{Sec:Scapula}

The bone remodelling theory, validated in the previous section, is applied to the intact scapula and to the scapula post procedure.
Prior to this, in order to elucidate features of the theory using a simpler geometry than the scapula, the implant is virtually implanted into a block. 
Finally, the full model of the scapula post procedure is developed.

\subsection{Response of the scapula prior to implant}
\label{SubSec:Scapula}

A geometry of the scapula produced by \citet{Mutsvangwa2015} was used.
The geometry is a statistical shape model, developed from 70 magnetic resonance imaging (MRI) scans. 
The pre-processing package ANSA \citep{ANSA} and the hexablocking technique are used to mesh the geometry using hexahedral elements.

\begin{figure}[htb!]
	\centering
	\includegraphics[keepaspectratio=true,width=0.8\textwidth]{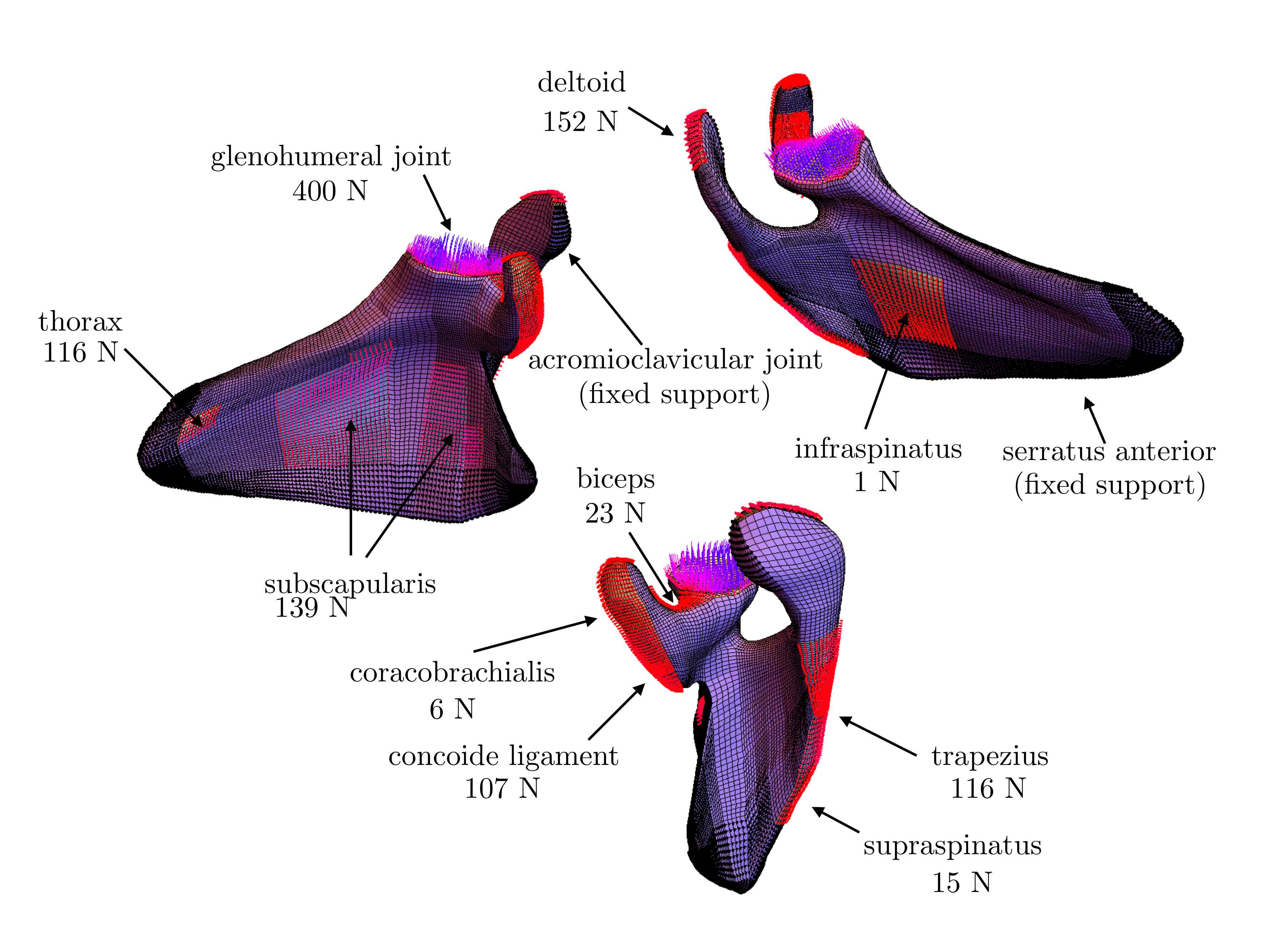}
\caption{The boundary conditions and geometry of the scapula model.}
   \label{fig:SSM_BCs}
\end{figure}

The boundary conditions were approximated from those reported by \citet{Sharma2013}, who applied the muscle forces at 90$^0$ abduction, as shown in Figure \ref{fig:SSM_BCs}. 
The levator scapula and rhomboid forces were omitted as these are relatively small. 
All forces were assumed to act normal to the surface.

The material properties for the scapula were chosen from various sources in the literature and are detailed in Table \ref{tab:ScapulaProperties}.
The mass conduction coefficient is chosen as $K_0 = 0.05 \text{ d mm}^{-2}$ to adjust the bounds of the density to fall within a physically reasonable range between $\rho_0 = 0 \text{ gcm}^{-3}$ and the density of cortical bone $ \rho_0 = 1.8 \text{ gcm}^{-3}$ as reported by \citet{Sharma2013}.

\begin{table}[h]
\centering
\begin{tabular}{l l l l}
\hline
\\
\textbf{Material Property} & \multicolumn{2}{l}{\textbf{Value}}  & \textbf{Source}\\
\\
\hline
\\
Young's modulus & $E$ & $= 193 \text{ N mm}^{-2}$  & \cite{Virani2008, ASTM_Glenoid} \\
Poisson's ratio & $\nu$ & $ = 0.25$ & \cite{Virani2008, ASTM_Glenoid} \\
Initial density & $\ini$ & $ = 0.6 \text{ g cm}^{-3}$ & \cite{Sharma2013} \\
Speed of adaptation & $c $ & $= 0.4 \text{ d mm}^{-2}$ & \cite{Waffenschmidt2012} \\
Attractor stimulus & $\Psi_0^*$ & $ = 0.01375 \text{ N mm}^{-2}$ & \cite{Waffenschmidt2012} \\
Mass conduction coefficient & $K_0 $ & $= 0.05 \text{ d mm}^{-2}$ &  \\
Time step size & $\Delta t $ & $ = 10$ d & \cite{Sharma2013} \\
\\
\hline
\end{tabular}
\caption{Material properties of the scapula}
\label{tab:ScapulaProperties}
\end{table}

\begin{figure}[htb!]
	\centering
	\includegraphics[keepaspectratio=true,width=1.0\textwidth]{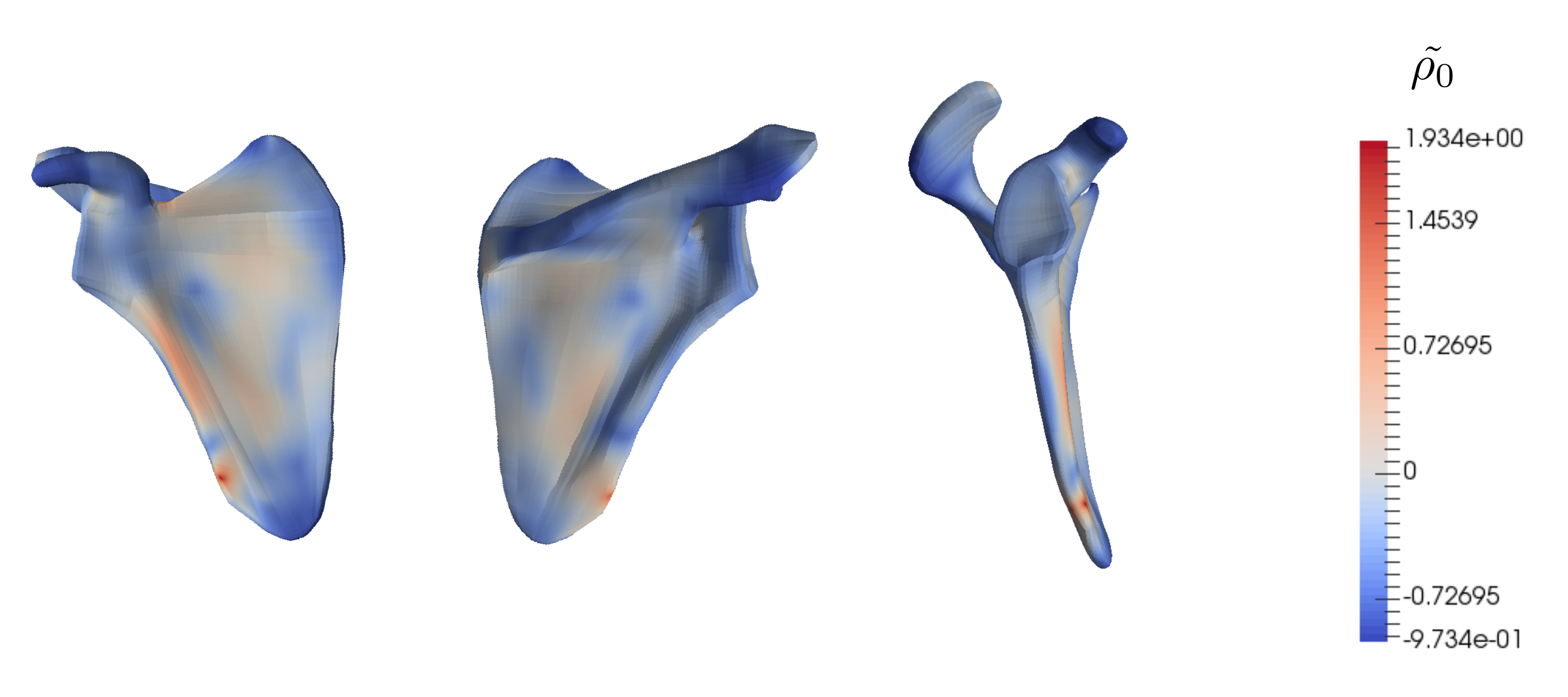}
\caption{The density distribution in the scapula with $\Delta t$ = 10 d at $t$ = 120 d.}
   \label{fig:SSM_Results}
\end{figure}

The predicted density distribution is shown in Figure \ref{fig:SSM_Results}.
The solution converges two time steps later than reported by \citet{Sharma2013} at $t =$ 120 d.
The resulting minimum and maximum relative density values of $\rel_{\text{min}} = $ -0.9734 and $\rel_{\text{max}} = $ 1.934 correspond to densities $\rho_{0} =  0.016 \text{ gcm}^{-3}$ and $\rho_{0} =  1.76 \text{ gcm}^{-3} $, respectively.
The density distribution is similar to that reported by \citet{Sharma2013}.
The discrepancies are explained by the differences in geometry, forces, muscle attachment points, material parameters and, most importantly, the different bone remodelling theory.

\subsection{Block example}
\begin{figure}[htb!]
	\centering
	\includegraphics[keepaspectratio=true,width=1.0\textwidth]{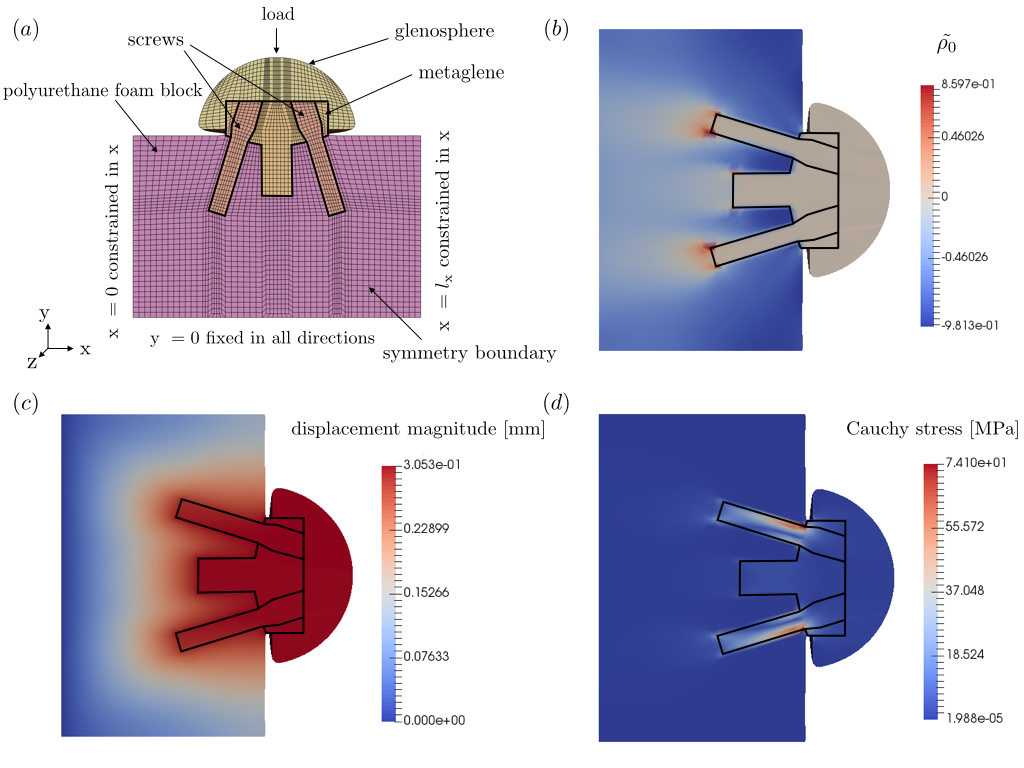}
\caption{The mesh of the block model, with all components and boundary conditions is shown in (a). The density, and norms of the displacement and Cauchy stress distribution at $t = 500$ d with $\Delta t = 10$ d and $K_0 = 0.01$ dmm$^{-2}$ are shown in (b), (c) and (d), respectively.}
   \label{fig:PU}
\end{figure}

This example investigates the remodelling response of a block that replicates the polyurethane foam block used in the ASTM F2028 \cite{ASTM_Glenoid} testing procedure to investigate glenoid component loosening.
However, the block now has the potential to remodel like bone.
A CAD model of the prosthesis was virtually implanted into the foam block. 
Owing to the symmetry of the problem, only half of the geometry was modelled and symmetry boundary conditions applied to the symmetry plane. 
The bottom edge was completely constrained and the left and right edges were constrained in the $x$-direction.
A compressive load of 85 \% of a standard body weight, which equals 750 N, was applied to the glenosphere. 
The mesh and boundary conditions are shown in Figure \ref{fig:PU} (a).

The same material properties as prescribed for the scapula were used for the block (see Table~\ref{tab:ScapulaProperties}).
The glenosphere, screws and metaglene were modelled as linear elastic materials with a Young's modulus of $E = 230\text{ } 000 \text{ N mm}^{-2}$ for the glenosphere and $E = 117 \text{ } 000 \text{ N mm}^{-2}$ for the screws and metaglene. 
The Poisson's ratio of the three linear elastic components was set to $\nu = 0.3$.
All components were modelled as perfectly bonded. 
The assumption that the density is continuous across the bone-screw interface leads to the artificial smoothing of the the density and related parameters as shown in Figure \ref{fig:PU}.
This could be addressed by double defining the density degrees of freedom at the interface.

The density, and norms of the displacement and Cauchy stress $\boldsymbol{\sigma} := J^{-1}\mathbf{P}\mathbf{F}^T$ distributions at $t = 500$ d with $\Delta t = 10$ d and $K_0 = 0.01 \text{dmm}^{-2}$ are shown in Figure \ref{fig:PU} (b), (c) and (d), respectively.
As expected, the stress is concentrated in the screws as these are stiff, load bearing components. 
As a result, the block's density increases in the vicinity of the screws.
As the greater part of the block does not experience any loading, the density decreases in these areas.  
It is  clear from the results that living bone would not evolve to this initial non-optimal cubic shape. 
Because the density is coupled to the material properties through the free energy, the bone is less stiff in the regions of low density and therefore offers little resistance to deformation. 

The influence of the time step size and mass conduction coefficient on the convergence rate of the problem is now investigated. 
Increasing the conduction coefficient, tightens the bounds of the predicted density (i.e.\ the difference between the maximum and minimum density values throughout the domain) and also increases the rate of convergence to an equilibrium state. 
As shown for the benchmark problem in Figure \ref{fig:KS_Fig2} b, changing the time step size results in the same equilibrium state, however, the time to equilibrium is different.
The time to equilibrium therefore has no physical interpretation. 

\subsection{Response of the scapula post reverse shoulder procedure}
\begin{figure}[htb!]
	\centering
	\includegraphics[keepaspectratio=true,width=1.0\textwidth]{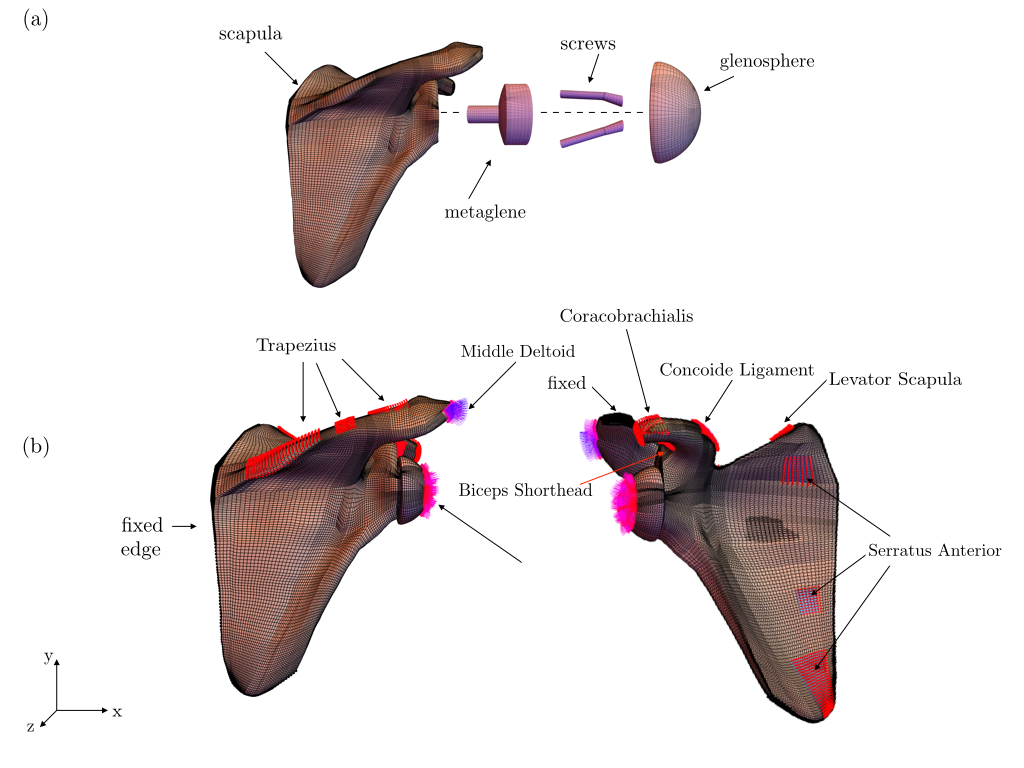}
\caption{The components of the reverse shoulder are shown in (a) and the boundary conditions are shown in (b).}
   \label{fig:RTSA_Components}
\end{figure}

The scapula post reverse total shoulder arthroplasty is now investigated. 
The statistical shape model of the scapula is the one used in Section~\ref{SubSec:Scapula}.
The reverse shoulder prosthesis is virtually implanted in the scapula, using the same implant geometry as for the block example in the previous section.
The surgical guidelines defined in the DePuy Delta XTEND reverse shoulder system's surgical techniques \citep{DeltaXtend} were followed and the virtual operation was performed in the software Mimics \citep{Mimics}.
The various components of the model are shown in Figure \ref{fig:RTSA_Components} (a).

The magnitudes of the applied muscle forces were obtained from three patients whose shoulders were simulated using the Newcastle Shoulder Model \citep{Kontaxis2009} in OpenSim \citep{OpenSim} - an open source musculoskeletal  modeling software.
The forces were also provided at 90$^0$ abduction, which is the same movement used for the intact scapula in Section~\ref{SubSec:Scapula}.
However, for the case of the reverse shoulder, the rotator cuff muscles (infraspinatus, supraspinatus, teres minor and subscapularis) are absent. 

The Newcastle Shoulder Model outputs point-wise muscle attachment locations and forces for each muscle group. 
The point-wise forces for each muscle group are averaged per patient, and each of the muscle forces are averaged over the three patients. 
The muscle attachment regions were estimated from the attachment points in the Newcastle Shoulder Model, as well as from the BioDigitalHuman project \citep{BioDigitalHuman}.
The forces were again assumed to act normal to the surface.
The muscle attachment regions and forces, as well as the fixed supports for the simulation are shown in Figure \ref{fig:RTSA_Components} (b). 

\begin{figure}[htb!]
	\centering
	\includegraphics[keepaspectratio=true,width=1.0\textwidth]{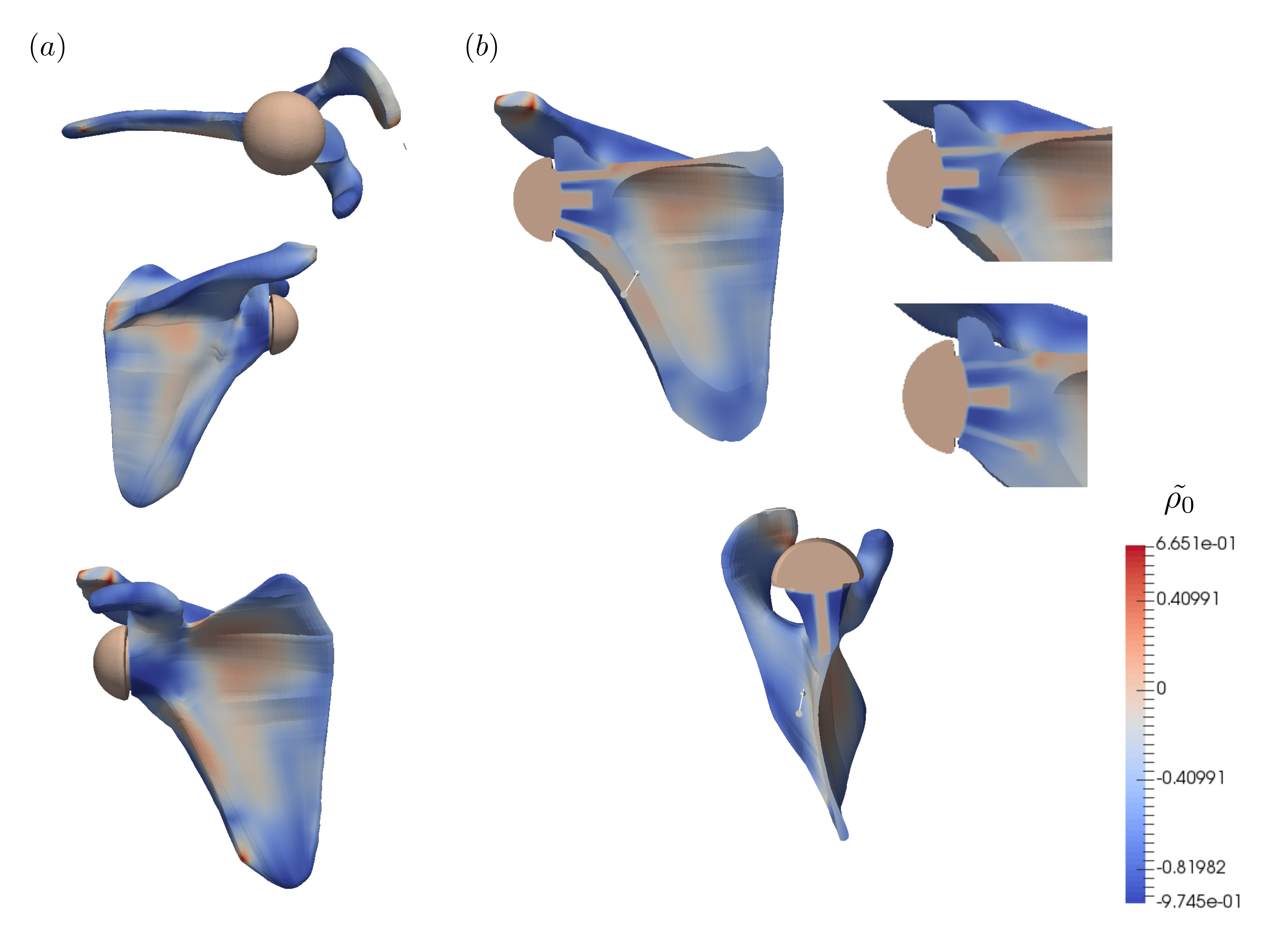}
\caption{The density distribution in the reverse shoulder after $\Delta t$ = 10 d, at $t$ = 120 d. Three views of the scapula are shown in (a) and section views are shown in (b).}
   \label{fig:RTSA_Results}
\end{figure}

The same material properties as for the intact scapula and the implant components in the block example were used.
The time step size $\Delta t = 10 $~d and equilibrium was reached at $t = 120 $~d.
The resulting density distributions are shown in Figure \ref{fig:RTSA_Results}.
The minimum and maximum relative density values of $\rel_{\text{min}} = -0.9745 $  and $\rel_{\text{max}} = 0.6651 $  correspond to densities $\rho_{0} = 0.0153 \text{ g cm}^{-3}$ and $\rho_{0} = 0.999 \text{ g cm}^{-3}$,   respectively. 
The maximum density is significantly less than for the intact scapula.
Diminished muscle forces act on the scapula, and the glenohumeral joint contact force, which is the largest force applied on the system, is applied directly onto the prosthesis. 
This results in the stresses concentrating in the screws, as observed in the block example.
As the screws are bearing the load, the stress in the bone is decreased, which results in a decrease in density.
This is a known phenomenon in orthopaedics, termed stress shielding.

The density distribution in the vicinity of the implant is comparable to the results from the block example.
In the regions where the muscle forces are applied, the density increases, as observed for  the intact scapula.
The density falls within the physiologically reasonable range. 

It is of concern that the density decreases significantly following the reverse shoulder procedure in regions directly below and adjacent to the metaglene, as density loss here can lead to implant loosening and ultimately to the failure of the prosthesis, resulting in devastating consequences for the patient.

\section{Discussion and conclusion}
\label{Sec:DiscussionConclusion}


An open-system model for bone remodelling developed by \citet{Kuhl2003a} was successfully applied to the case of a scapula post reverse shoulder arthroplasty. 

The remodelling code developed was validated against a one-dimensional benchmark problem.
Different aspects of the theory were investigated. 
It was shown that the attractor stimulus has to be small enough to ensure density apposition. 
The mass conduction coefficient $K_0$ should be calibrated to adjust the bounds of the density.
The mass conduction coefficient and time step size were also shown to have an influence on the rate of convergence.

The validated model was applied to a statistical shape model of the scapula. 
The forces at 90$^0$ abduction were taken from literature and applied normal to the surface.
It is difficult to compare the model to the literature as the geometry, muscle attachment points and even the bone remodelling theory are significantly different. 
Nevertheless, the results are reasonable once the mass conduction coefficient is calibrated so that physical density bounds are predicted. 
In most other remodelling theories, the density bounds are explicitly enforced, which is not necessary in the current model.

The shoulder was then examined post-procedure. 
First, the implant was placed in a block, which was modelled as bone, followed by an investigation of the scapula post procedure.  
The resulting density values in the scapula post procedure were significantly smaller than for the intact scapula, due to the absent rotator cuff muscles, and because the screws are the load bearing components in the system.
The response to loading results in a densification around the screws and density resorption underneath and adjacent to the implant, which is a cause for concern. 
The areas where the muscle forces act, experience density apposition.
A definite conclusion on the remodelling response can however not be drawn, as many uncertainties are present and only one loading scenario is considered.

In summary, the remodelling model provides physically reasonable and justifiable results.
The density bounds are enforced naturally by adjusting the mass conduction coefficient to suit the problem.
What is outstanding, is a comprehensive study of the scapula and calibration of the model parameters against experimental studies.
Further, the exact geometry has to be imported to the muscoloskeletal shoulder model and a number of everyday loading scenarios should be considered.
An extension of the model to include anisotropy considerations in the density evolution will be important as the density distribution in the scapula is anisotropic.
Further, modelling of the contact between the implant and bone may yield a more predictive model.
Once these extensions have been realised, the model provided here could be used as the basis for a reliable and powerful predictive tool in the research and design process of reverse shoulder arthroplasty.




\section*{Acknowledgements}
The help and information for the reverse shoulder model provided by Jonathan Glenday from the Division of Biomedical Engineering at UCT and Andreas Kontaxis at the Leon Root Motion Analysis Laboratory at the Hospital for Special Surgery in New York City is greatly appreciated. 
Further, the statistical shape model provided by Tinashe Mutsvangwa in the Department of Biomedical Engineering at UCT is gratefully acknowledged.

H.L. and A.M. acknowledge funding from the National Research Foundation through the South African Research Chair in Computational Mechanics (SARCCM, grant agreement SARChI-CM- 47584) and the funding from the University of Cape Town through the Master's Research Scholarship.
The financial assistance of the National Research Foundation (NRF) through the Scarce Skills Master's Scholarship (NRF Grant Number 102021) towards this research is hereby acknowledged. 
Opinions expressed and conclusions arrived at, are those of the authors and are not necessarily to be attributed to the NRF.

\newpage
\bibliographystyle{model1-num-names}
\bibliography{bibs_paper.bib}





 \end{document}